\documentclass[5p]{elsarticle}

\usepackage{amsmath}
\usepackage{feynmf}
\usepackage{latexsym}
\usepackage{amssymb}
\usepackage{graphicx}
\usepackage{rotating}
\usepackage{latexsym}
\usepackage{verbatim}
\usepackage{lineno}

\begin{document}
\linenumbers

 \def\be{\begin{equation}}
 \def\ee{\end{equation}}
 \def\l{\lambda}
 \def\a{\alpha}
 \def\b{\beta}
 \def\g{\gamma}
 \def\d{\delta}
 \def\e{\epsilon}
 \def\m{\mu}
 \def\n{\nu}
 \def\t{\tau}
 \def\p{\partial}
 \def\s{\sigma}
 \def\r{\rho}
 \def\sl{\ds}
 \def\ds#1{#1\kern-1ex\hbox{/}}
 \def\sla{\raise.15ex\hbox{$/$}\kern-.57em}
 \def\nn{\nonumber}
 \def\bea{\begin{eqnarray}}
 \def\eea{\end{eqnarray}}
 \newcommand{\bth}{{\bf 3}}
 \newcommand{\btw}{{\bf 2}}
 \newcommand{\bon}{{\bf 1}}
 \def\QQ{{Q_Q}}
 \def\QU{{Q_{U^c}}}
 \def\QD{{Q_{D^c}}}
 \def\QL{{Q_L}}
 \def\QE{{Q_{E^c}}}
 \def\QHu{{Q_{H_u}}}
 \def\QHd{{Q_{H_d}}}
 \def\cA{\mathcal{A}}
 \def\Tr{\textnormal{Tr}}
 \def\th{\theta}
 \def\pd{\partial}
 \def\thth{\theta^2 \bar\theta^2}
 \def\sb{\bar{\s}}
 \def\psib{{\bar{\psi}}}
 \def\f{\phi}
 \def\Eps{\epsilon^{\mu\nu\rho\sigma}}
 \def\({\left(}
 \def\){\right)}
 \def\[{\left[}
 \def\]{\right]}
 \def\lb{{\bar\l}}
 \def\half{\frac{1}{2}}
 \def\la{\langle}
 \def\ra{\rangle}
 \def\numeq{n^{eq}}
 \def\D{\Delta}
 \def\cA{\mathcal{A}}
 \def\cw{\cos\th_W}
 \def\sw{\sin\th_W}
 \def\cscw{\text{cosec}\th_W}
 \def\cotw{\cot\th_W}
 \def\numeq{n^\text{eq}}
 \def\slashed{\ds}
 \def\sl{\ds}
 \def\ds#1{#1\kern-1ex\hbox{/}}
 \def\sla{\raise.15ex\hbox{$/$}\kern-.57em}
 \def\Stuckelberg{St\"uckelberg }



\begin{frontmatter}

\title{Anomalous $U(1)'$ and Dark Matter}

\author[label1]{Andrea
Mammarella}

\address[label1]{Dipartimento di Fisica dell'Universit\`a di Roma ,``Tor Vergata" and
I.N.F.N.~ -~ Sezione di Roma ~ ``Tor Vergata'',\\
 Via della Ricerca  Scientifica, 1 - 00133 ~ Roma,~ ITALY}

\begin{abstract}
We have studied the lightest masses in the fermionic sector of an anomalous $U(1)'$ extension of the 
Minimal Supersymmetric Standard Model (MSSM) inspired by brane constructions. The LSP of this model 
is an XWIMP (extremely weak interaction particle). We have studied its relic density in the cases in which
there is mixing with the neutralinos of the MSSM and in the case in which there is not such mixing.  We 
have showed that this extended model can satisfy the WMAP data. To perform
this calculation we have modified the DarkSUSY package.

\end{abstract}

\end{frontmatter}

\section{Introduction}

There has been much work recently to conceive an intersecting brane model with the gauge and matter 
content of the Standard Model (SM) of particle physics \cite{Marchesano:2007de,Blumenhagen:2005mu,Lust:2004ks,Kiritsis:2003mc}.
One of the features of such models is the presence of extra anomalous gauge $U(1)'$s
whose anomaly is cancelled via the Green-Schwarz mechanism. The anomalies of these models
 are cancelled without assumption on the newly introduced quantum numbers. One model
of this type have been introduced and discussed in \cite{Anastasopoulos:2008jt} and \cite{Fucito:2010dj}. We have studied 
the compatibility of this model with the WMAP data 
\cite{FLMR,FLM} (see also \cite{Coriano:2010ww,Coriano:2010ws} for related work). In this model we have
two new contribution to the neutralinos mass matrix:
one coming from the superpartner of the St\"uckelberg boson (St\"uckelino) and the other from the superpartner 
of the gauge boson mediating the extra $U(1)'$ (Primeino). By taking some simplifying and reasonable 
assumptions on the fermion masses entering the
soft supersymmetry breaking lagrangian, the LSP turns out to be mostly a mixture of St\"uckelino and Primeino. 
It is also easy to realize that, given the 
simplifying assumptions mentioned above, the LSP is interacting with the MSSM particles with a coupling suppressed by 
the inverse of the mass of the extra gauge boson of the theory which must be at least of the order of the TeV 
for phenomenological reasons. 
The LSP is then an XWIMP (Extra Weakly Interacting Massive Particle), a class
of particles which have already been studied in literature \cite{Kors:2005uz,Feldman:2006wd}.
The cross section of these LSPs is too weak to give the right relic abundance. This is why one has to resort 
to coannihilations with NLSPs. In our case, the cross section for the annihilations of the LSP with the NLSP 
and that for the coannihilations of the two species differ for some orders of magnitude. Once again this situation 
is not new in literature \cite{Klein:1999im,reliccoann,Edsjo:2003us}
but needs to be treated carefully: the two species will not decouple as far as there will be some MSSM 
particles to keep them in equilibrium. Moreover these particles must be relativistic so that their abundance 
is enough to foster the reaction.
There are two cases. First, the anomalous sector of the neutralinos mass matrix can be decoupled from the MSSM
sector: this implies that the LSP can be either pure anomalous (St\"uckelino-Primeino mix) or pure MSSM. This is
called the no mixing case and we have studied it in the case in which the LSP is pure anomalous and coannihilates
with a pure MSSM NLSP. The second case, called the mixing case, is that in which there are mixing terms between 
the two sectors of the neutralinos mass matrix. So we can have LSP and NLSP that can be a general mixing of the six gauge
eigenstates of the theory. We will show that also in this general case the mixing is constrained and then we will
study the case in wich the LSP is mostly anomalous with fractions of MSSM eigenstates, while the NLSP is 
mostly of MSSM origin whit small fractions of anomalous eigenstates.\\
In the subsequent sections we will describe the results for the relic density in both cases. We have obtained that there are
 regions of the parameter space  in which the WMAP data can be satisfied in both cases. These results
are obtained using a modified version of the DarkSUSY package which takes into account the couplings of our model.

\section{Model \label{sec2}}
Our model \cite{Anastasopoulos:2008jt} is an extension of the MSSM with an extra $U(1)$. 
The charges of the matter fields with respect to the symmetry groups are given in table 1.

  \begin{table}[h]
  \centering
  \begin{tabular}[h]{|c|c|c|c|c|}
   \hline & SU(3)$_c$ & SU(2)$_L$  & U(1)$_Y$ & ~U(1)$^{\prime}~$\\
   \hline $Q_i$   & $\bth$       &  $\btw$       &  $1/6$   & $Q_{Q}$ \\
   \hline $U^c_i$   & $\bar \bth$  &  $\bon$       &  $-2/3$  & $Q_{U^c}$
\\
   \hline $D^c_i$   & $\bar \bth$  &  $\bon$       &  $1/3$   & $Q_{D^c}$
\\
   \hline $L_i$   & $\bon$       &  $\btw$       &  $-1/2$  & $Q_{L}$ \\
   \hline $E^c_i$   & $\bon$       &  $\bon$       &  $1$     &
$Q_{E^c}$\\
   \hline $H_u$ & $\bon$       &  $\btw$       &  $1/2$   & $Q_{H_u}$\\
   \hline $H_d$ & $\bon$       &  $\btw$       &  $-1/2$  & $Q_{H_d}$ \\
   \hline
  \end{tabular}
  \caption{Charge assignment.}\label{QTable}
  \end{table} Gauge invariance imposes constraints that leave independent only three of these
charges. In this model we have chosen: $Q_{Q},~Q_L$ and $Q_{H_u}$. The anomalies
induced by this extension are cancelled by the GS mechanism. 
Each anomalous triangle diagram is parametrized by a coefficient $b_2^{(a)}$ (entering the lagrangian)
with the assigment:

\bea
   \cA^{(0)}:     &&\ U(1)'-U(1)'-U(1)' \rightarrow b_2^{(0)} \\
   \cA^{(1)}: &&\ U(1)'-U(1)_Y - U(1)_Y \rightarrow b_2^{(1)}\\
   \cA^{(2)}:     &&\ U(1)'-SU(2)-SU(2) \rightarrow b_2^{(2)}\\
   \cA^{(3)}:     &&\ U(1)'-SU(3)-SU(3) \rightarrow b_2^{(3)}\\
   \cA^{(4)}:    &&\ U(1)'-U(1)'-U(1)_Y \rightarrow b_2^{(4)}
\eea The mass of the extra boson is parametrized by 
$M_{V^{(0)}}=4 b_3 g_0\sim~1~TeV$, where $g_0$ is the coupling of 
the extra $U(1)'$. The terms of the Lagrangian that will
contribute to our calculation are \cite{Anastasopoulos:2008jt, FLMR}:

\bea
 &&\L_{\text{St\"uckelino}}={\frac{i}{4}} \psi_S \s^\m \pd_\m \psib_S -\sqrt2 b_3 \psi_S
\l^{(0)} \nn \\ 
   &&-{\frac{i}{2\sqrt2}} \sum_{a=0}^2b^{(a)}_2  \Tr \( \l^{(a)} \s^\m
\sb^\n F_{\m
\n}^{(a)} \) \psi_S \\
   &&-{\frac{i}{2\sqrt2}} b^{(4)}_2 \[ {\frac{1}{2}} \l^{(1)} \s^\m \sb^\n
F_{\m
\n}^{(0)} \psi_S
   + (0 \leftrightarrow 1) \] +h.c.
\label{axinolagr} \nn \eea The parameters $b_2^{(i)}$ are inversely proportional
to $M_{V^{(0)}}$, so they are much smaller than the couplings of the SM. This
implies that the primeino and the st\"{u}ckelino are XWIMPS.\\

The neutral mixing matrix is:
     
\be
      \( \begin{array}{c} B^{\m}\\
                          W^{3 \m} \\ C^{\m} \end{array} \) = M
            \( \begin{array}{c} A^{\m}\\
                                Z_0^{\m}\\ Z'^{\m} \end{array} \)
\label{neumix}  \ee Defining $\emph{an} \equiv g_0 ~ Q_{H_u} \frac{2 v^2}{2 M^2_{Z'}}$ we have at tree level:
\be  M= \( \begin{array}{ccc} c_W & -s_W & s_W \sqrt{g_1^2 + g_2^2}~an \\
s_W & c_W  & c_W \sqrt{g_1^2 + g_2^2}~an\\
  0 & c_W g_2+ s_W ~g_1)~an & 1    \end{array} \)
\label{neumass} \ee where $c_W$ is $cos(\theta_W)$, $s_W$ is $sin(\theta_W)$. $g_1,~g_2$ are the couplings
of the SM electro-weak $SU(2) \times U(1)$ group. The structure of this matrix leaves the electromagnetic
sector and the related quantum numbers unchanged with respect to the MSSM ones.
We can see from eq. (\ref{neumass}) that the case of mixing between the anomalous sector and the MSSM 
sector is that in which $an\ne 0$, while the case of no mixing is that in which $an =0$. From the definition of
$an$ it's clear that $an=0 \Leftrightarrow Q_{H_u}=0$.\\
Last, we remember \cite{Anastasopoulos:2008jt,FLMR} the general form of the neutralinos mass matrix at 
tree level:
\be
    {\bf M}_{\tilde N}
     =   \(\begin{array}{cccccc}
           M_S & \sqrt{2} \frac{M_{Z'}}{2} & 0 &0 & 0 & 0 \\
           \dots & M_0 & 0  & 0 & - g_0 v_d Q_{H_u}  & g_0 v_u Q_{H_u}   \\
           \dots & \dots & M_1 & 0 & -\frac{g_1 v_d}{2} & \frac{g_1 v_u}{2} \\
           \dots & \dots& \dots & M_2 & \frac{g_2 v_d}{2} & -\frac{g_2 v_u}{2} \\
           \dots & \dots & \dots & \dots & 0 & -\m  \\
          \dots & \dots & \dots & \dots & \dots & 0
         \end{array}\) \nn ~~~~
\label{massneu}\ee
where $M_S$ and $M_0$ are the soft masses of the st\"{u}ckelino
and of the primeino, respectively and $v_u,~v_d$ are the vevs (vacuum expectation values) of the Higgs fields.
We can see that in case of no mixing ($Q_{H_u}=0$), there are no terms different
from $0$ except for the two diagonal boxes that refer to the anomalous and the MSSM
sectors.

\section{Boltzmann Equation and coannihilations}
The relic density of a certain particle is given by the Boltzmann Equation (BE) \cite{Gondolo}:
\be \frac{dn}{dt}=-3Hn-\sum_{ij}\langle \s_{ij} v_{ij} \rangle
(n_i n_j - n^{eq}_i n^{eq}_j) \label{boltz3} \ee $\s_{ij}$ is the annihilation cross section:

\be X_i X_j \rightarrow f\bar{f} \ee $v_{ij}$ is the absolute value
of relative velocity between i-th and j-th particle, defined by:
$v_{ij}=|v_i-v_j|$ , $n_i$ is the number of particle per unit of
volume of the i-th specie (index 1 refers to LSP),
 $n_i^{eq}$ is the number of particle per unit of
volume of the i-th specie in thermal equilibrium. $n=\sum_i n_i$,
H is the Hubble constant and $v$ is the relative velocity of the 
initial particles. \newline
If the LSP is an XWIMP, it is known \cite{Kors:2005uz} that its relic density does not
satisfy the WMAP observations \cite{WMAP}. So, in order to achieve the sperimental 
results, we need coannihilations. This phenomenon occurs when there is at least
one particle with mass of the order of the mass of the XWIMP LSP. If the interaction of the
coannihilating particles are strong enough they can lower the relic density of the
LSP by several orders of magnitude and so we can obtain results in agreement with the 
sperimental data.\\
Although a general tractation can be found in \cite{tesiandrea}, for our purposes we are
interested only in the cases in which the number $N$ of the coannihilating particles is
$2$ or $3$. In the case $N=2$ we have the following thermal averaged cross section:

\bea
 &&\langle \s_{eff}^{(2)} v \rangle = \langle\sigma_{11}v\rangle (\frac{n_1^{eq}}{n^{eq}})^2+\nn \\
&&2 \langle\sigma_{12}v\rangle \frac{n_1^{eq}n_2^{eq}}{(n^{eq})^2}+ 
\langle\sigma_{22}v\rangle (\frac{n_2^{eq}}{n^{eq}})^2=  \\
&&\langle\sigma_{22}v\rangle\frac{\langle\sigma_{11}v\rangle/\langle\sigma_{22}v\rangle
+2\langle\sigma_{12}v\rangle/\langle\sigma_{22}v\rangle Q+Q^2}{(1+Q)^2} \nn
\label{sigma2eff}
\eea where $Q=n_2^{eq}/n_1^{eq}$. The first term in the numerator can
be neglected because the St\"uckelino annihilation cross section
is suppressed by a factor $(b_2^{(a)})^4$ with respect to the MSSM
neutralino annihilations (see the previous section) and thus
$\langle\sigma_{11}v\rangle\ll \langle\sigma_{22}v\rangle$.\\
With the same assumptions of the case $N=2$ the formula for the case $N=3$
is:
\bea
 \la \s_\text{eff}^{(3)} v \ra
               &\simeq& \frac{\la \s_{22} v \ra  Q_2^2 + 2 \la \s_{23} v \ra  Q_2 Q_3 + \la \s_{33} v \ra  Q_3^2}{\(1+Q_2+Q_3\)^2}
\eea with 
\be
Q_i=\frac{n_i^{eq*}}{n_1^{eq}}
\ee This expressions show that the factors $Q_i$ can raise/lower the thermal averaged 
cross section and thus modify the related relic density. However to obtain the result
for the relic density it is necessary to numerically solve the BE. To achieve this we have used 
the DarkSUSY package, adding the interactions introduced by our extended model. 
In the following section we describe our results.

\section{Numerical computations and results \label{sec3}}
To perform a numerical analysis on the extended model we have modified the DarkSUSY
package adding new fields and interactions
introduced by the anomalous extension. The free parameters that we use in our numerical simulations are
the seven ones used in the MSSM-7 model: the $\m$ mass, the
wino soft mass $M_2$, the parity-odd Higgs mass $M_{A_0}$,
$tg \b$, the sfermion mass scale $m_{sf}$, the two Yukawas
$a_t$ and $a_b$. We add to this set five parameters which define 
the $U(1)'$ extension: the st\"{u}ckelino soft mass $M_S$, 
the primeino soft mass $M_0$, the $U(1)'$ charges $Q_{H_u}$, $Q_Q$, 
$Q_L$.\\
\subsection{No-mixing case}
\begin{figure}
 \includegraphics[scale=0.65]{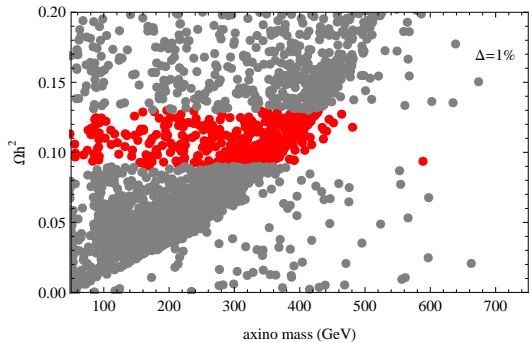}
\caption{Relic density for a bino-higgsino type NLSP and mass gap 1 \%}
\label{nomixb1}
\end{figure}

\begin{figure}
 \includegraphics[scale=0.65]{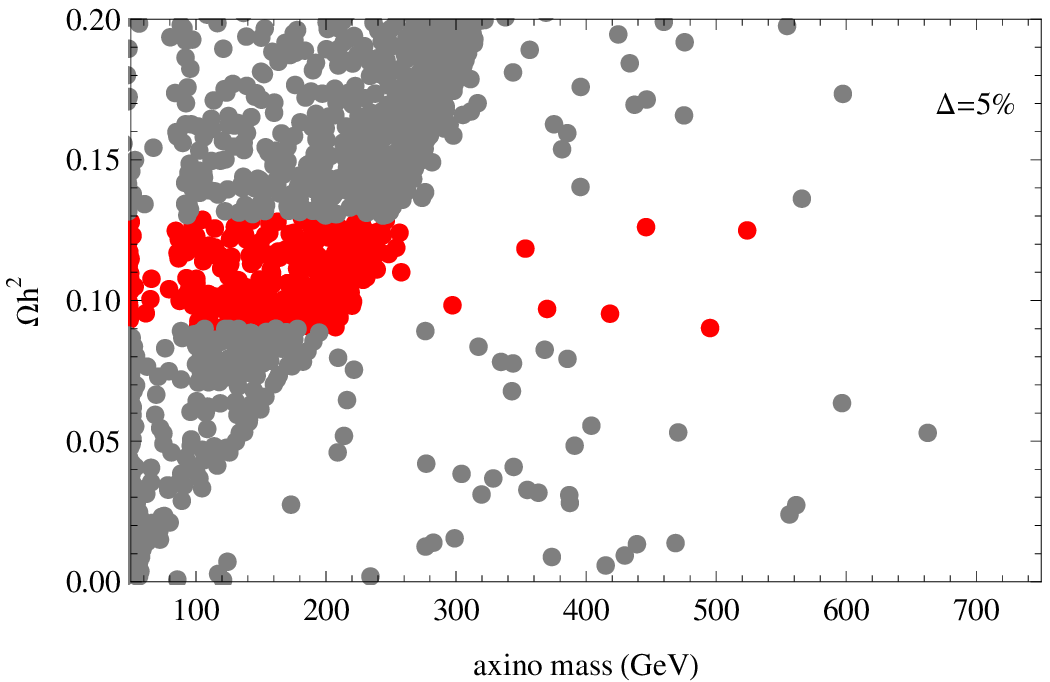}
\caption{Relic density for a bino-higgsino type NLSP and mass gap 5\% }
\label{nomixb2}
\end{figure}

\begin{figure}
 \includegraphics[scale=0.65]{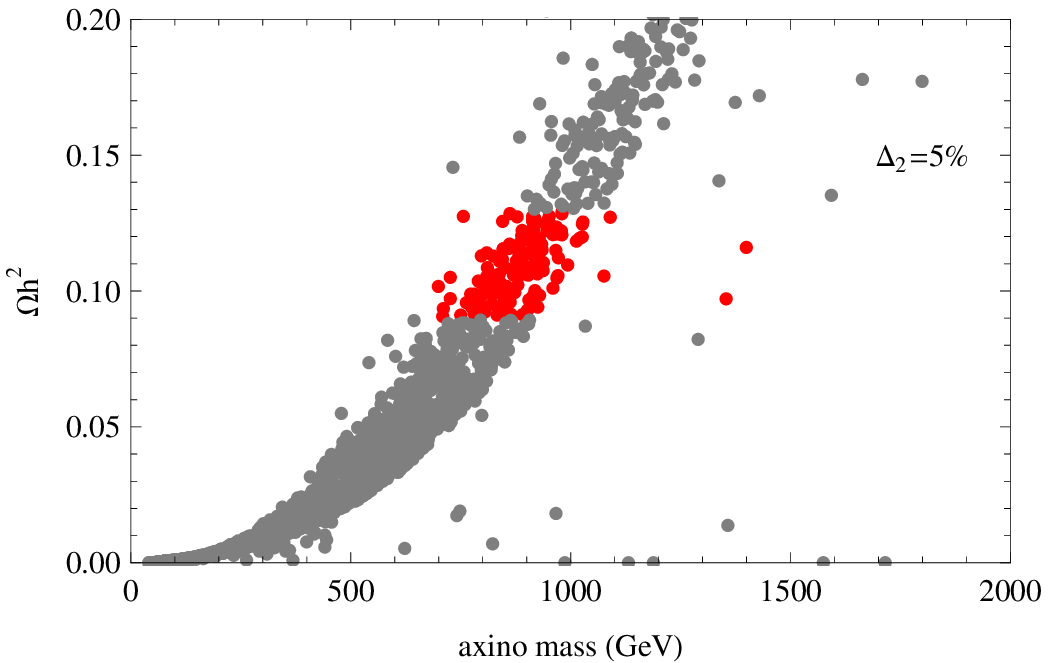}
\caption{Relic density for a wino type NLSP and mass gap 5\%} 
\label{nomixw1}
\end{figure}

\begin{figure}
 \includegraphics[scale=0.65]{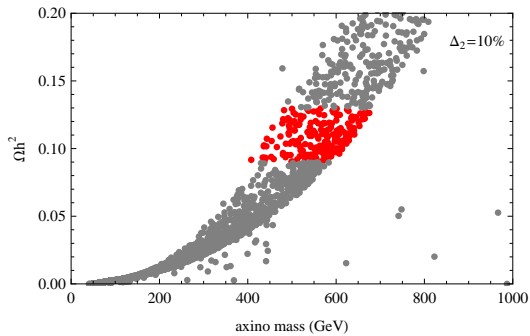}
\caption{Relic density for a wino type NLSP and mass gap 10 \%} 
\label{nomixw2}
\end{figure}

As we have seen if there is no-mixing we have to impose $Q_{H_u}=0$. From eq. (\ref{massneu}), this implies
that the LSP can have either pure anomalous origin or pure MSSM origin. We are interested in the
first case. Furthermore, we assume for simplicity that the LSP is a pure  st\"{u}ckelino. Because
in this case the anomalous and the MSSM sectors are decoupled, we can choose the model parameters
to keep fixed the mass gap between the LSP and the NLSP. So we have studied the behaviour of
the model in function of the mass of the LSP for a defined mass gap.\\
We have studied separately the two cases in which the NLSP is mostly a bino and in which the NLSP is mostly a
wino. In the first case the coannihilation involves only the LSP and the NLSP while in the latter case it involves also
the lightest chargino, that in the MSSM is almost degenerate in mass with the wino. So they are different situations.
The results of the relic density calculations are showed in figures \ref{nomixb1} and \ref{nomixb2} for the bino NLSP case and in
figures \ref{nomixw1} and  \ref{nomixw2} for the wino LSP case.

In those figures we have shown the most significant results, but we have found model that satisfy WMAP data
up to $10\%$ mass gap for the bino NLSP case and up to $20\%$ for the wino NLSP case.

\subsection{Mixing case}
No mixing implies that $Q_{H_u}\ne 0$, so
we can't have control on the mass gap between the LSP and NLSP,
because the mass matrix is no more decoupled in two blocks. So we
choose to let all parameters unconstrained and therefore to
collect data in the mass gap ranges 
$0\div5\%$, $5\div10\%$.
In each case we have started our study scanning the parameter space
in search of the region permitted by the experimental and theoretical constraints,
i.e. the region in which we could satisfy the WMAP data with a certain choice
of the model parameters.
After that we have numerically explored this regions to find sets of parameters that
satisfy the WMAP data for the relic density. We have found many suitable combinations
for both types of NLSP. So we have chosen some of these successful 
models and we have computed the relic density keeping constant all but two 
parameters and plotting the results.\\
We have found regions of the parameter space in which the WMAP data
are satisfied for mass gap over $20\%$, but in the following we will only show results for the regions
$0\div5\%$ and $5\div10\%$, because they are more significant. For simplicity
we will refer to these regions as \textquotedblleft
$5\%$ region \textquotedblright and \textquotedblleft $10\%$ region\textquotedblright
respectively.\\

\subsubsection{General results}
In this section we want to list some results that are valid for both types
of NLSP.\\ First of all, given the constraints on the 
neutral mixing described in \cite{Langacker:2008yv}, we have obtained that
$-1 \lesssim Q_{H_u} \lesssim 1$. This constraint was not appearing in our preliminary analysis in \cite{FLMR}.
This implies that also in our general case the 
mixing between our anomalous LSP and the NLSP is small.\\
We have checked that there are suitable parameter space regions in which the WMAP data are satisfied
and we have found that this is true for all possible composition of our anomalous LSP.\\
We have also checked that in each region we have studied there are no divergences or unstable behaviours in our 
numerical results.\\
We have verified that the relic density is strongly dependent
on the LSP and NLSP masses and composition while it is much less dependent on the
other variables. Anyway it can be shown that there are cases 
in which the parameters not related to the LSP or NLSP can play
an important role. We will show an example of this  case in a forthcoming subsection.

\subsubsection{Bino-higgsino NLSP}
If the NLSP is mostly a bino-higgsino we have a two particle coannihilation.\\
We chose two sample models which satisfy the WMAP data \cite{WMAP} with mass gaps $5 \%$ and $10 \%$.
We study these models to show the dependence of the relic density from the
LSP composition and from the mass gap. To obtain these result, we have performed
a numerical simulation in which we vary only the st\"{u}ckelino and the primeino soft masses.
The results are showed in figure 1, with the conventions:
\begin{itemize}
 \item Inside the continous lines we have the region in which \newline $(\Omega h^2)_{WMAP} \sim \Omega h^2$
 \item Inside the thick lines we have the region in which \newline $(\Omega h^2-3 \s)_{WMAP}<\Omega h^2<(\Omega h^2+3 \s)_{WMAP}$
 \item Inside the dashed lines we have the region in which \newline $(\Omega h^2-5 \s)_{WMAP}<\Omega h^2<(\Omega h^2+5 \s)_{WMAP}$
 \item Inside the dotted lines we have the region in which \newline $(\Omega h^2-10 \s)_{WMAP}<\Omega h^2<(\Omega h^2+10 \s)_{WMAP}$
\end{itemize}

\begin{figure}[h!]
\includegraphics[scale=0.65]{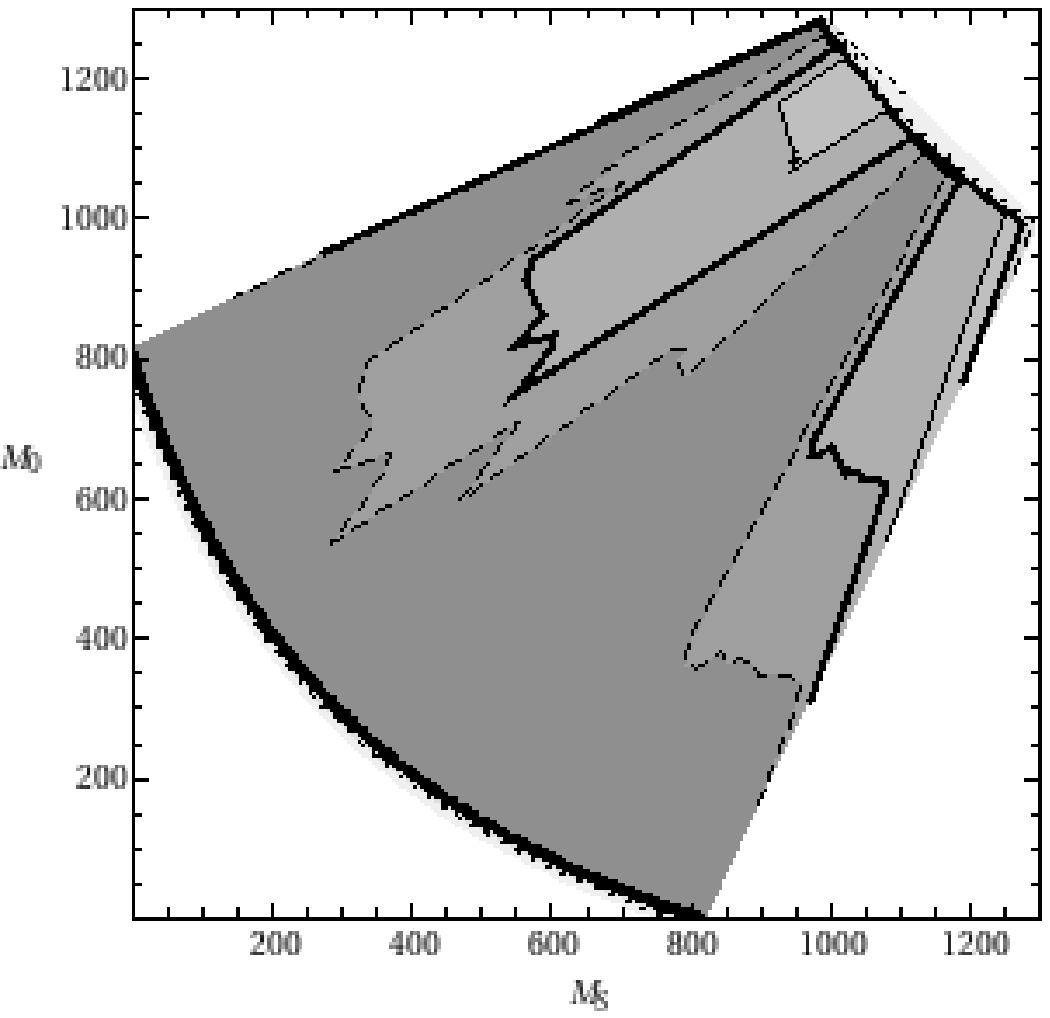}
\includegraphics[scale=0.5]{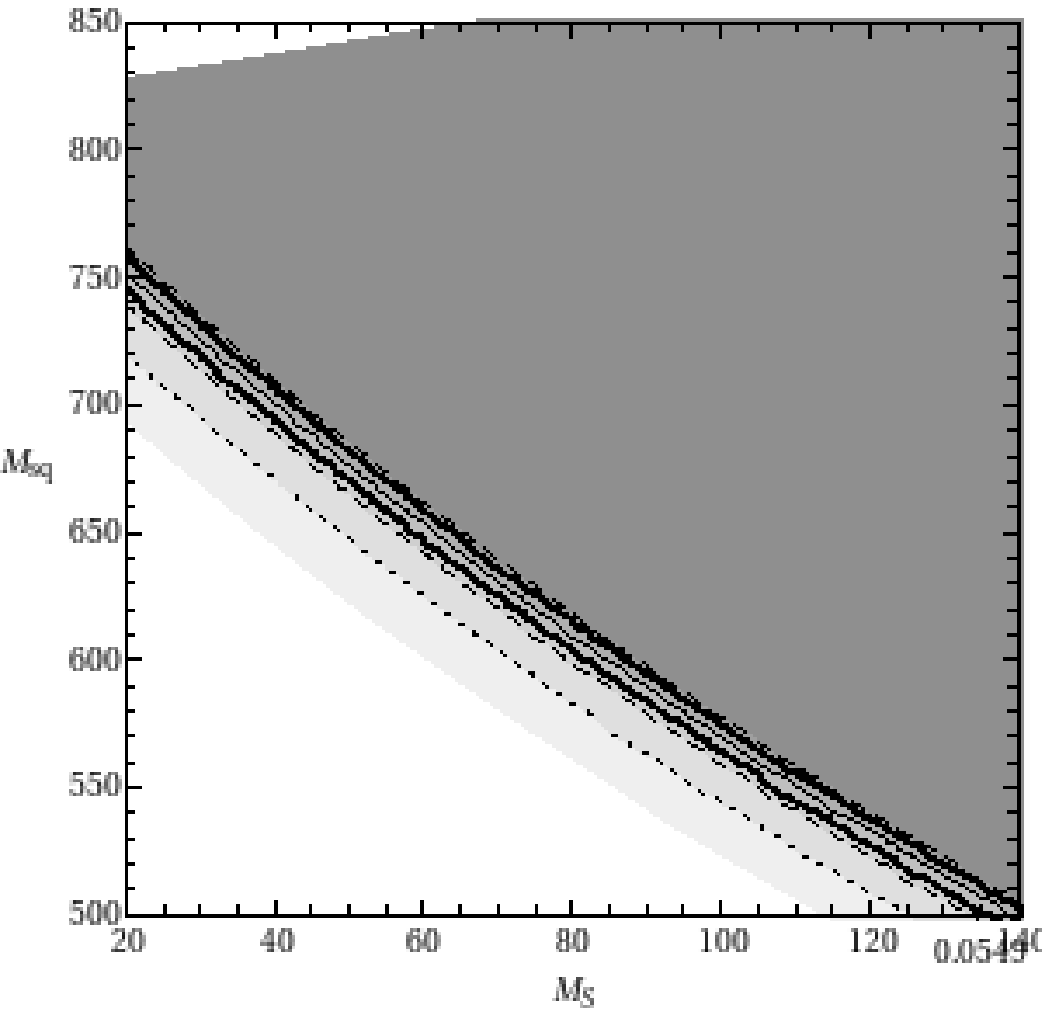}
\includegraphics[scale=0.8]{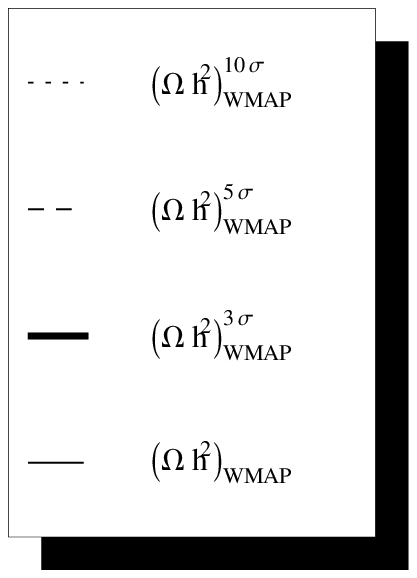}
\includegraphics[scale=0.65]{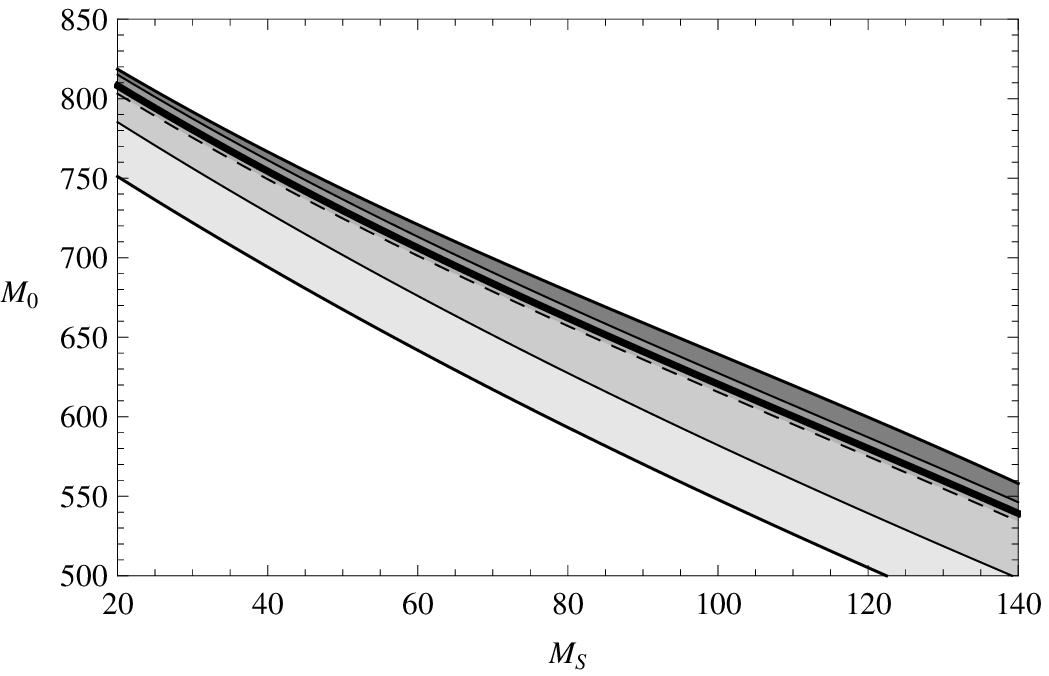}
\caption{Plot of the relic density of the LSP vs the st\"{u}ckelino
and primeino masses. The first plot shows the case of mass gap 5 $\%$. The second plot
is a zoom of the first in the region between 20-140 GeV for st\"{u}ckelino mass and 500-850 GeV
for primeino mass. The third plot shows the case of mass gap 10 $\%$ in the same region of the second plot.
All the masses are expressed in GeV}
\end{figure} Going from the region with a $5\%$ mass gap to that with a $10\%$
mass gap there is a large portion of the parameter space in which the 
WMAP data cannot be satisfied, while the regions showed in the second and third plot in
fig. 1 are similar and thus are mass-gap independent.

\subsubsection{Wino-higgsino NLSP}
If the NLSP is mostly a wino-higgsino we have a three particle coannihilation,
because the lightest wino is almost degenerate in mass with the lightest chargino,
so they both contribute to the coannihilations.\\
In this case we perform the same numerical calculation illustrated in the 
previous subsection. We have extensively studied a sample model with mass gap $10\%$, 
showing an example of funnel region, a resonance that occurs when $2~M_{LSP} \sim M_{A_0}$.
In our sample model $M_{LSP} \sim 330~GeV$, while $M_{A_0}\sim 630~GeV$ and this
leads to the relic density plot showed in figure 2, with the same conventions used in figure 1.
\begin{figure}[h!]
\includegraphics[scale=0.65]{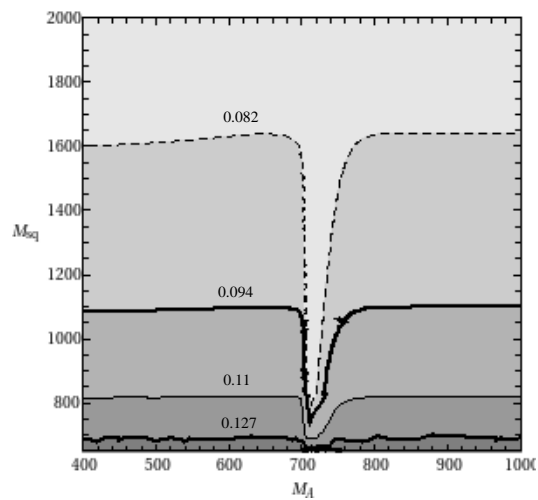} 
\caption{Example of a funnel region for a wino-like NLSP, whose mass gap with
the LSP is $10\%$}
\end{figure} So we can state that also in the case of anomalous LSP we can have a
behaviour similar to that of the MSSM, given that the LSP coannihilates with a MSSM NLSP.

\section{Conclusion \label{sec4}}
We have modified the DarkSUSY package in the routines which calculate the cross section of a given
supersymmetric particle (contained in the folder $\sim$/src/an) adding all the new interactions introduced
by our anomalous extension of the MSSM. We have also written new subroutines to calculate amplitudes that 
differ from those already contained in DarkSUSY. We have also modified the routines that generate the
supersymmetric model from the inputs, adding the parameters necessary to generate the MiAUSSM \cite{Anastasopoulos:2008jt}
and changing the routines that define the model (contained in $\sim$/src/su) accordingly. Finally
we have written a main program that lets the user choose if he wants to perform the relic density calculation
in the MSSM or in the MiAUSSM. The code of our version of the package is available contacting 
andrea.mammarella\newline @roma2.infn.it.\\
These modifications have permitted to extensively numerically explore the parameters space for an anomalous 
extension of the MSSM without restriction on the neutral mixing or on the free
parameters of the model. We have verified that our model does not lead to any divergence
or instability. \\ We have studied separately the case in which there is mixing between the anomalous and 
the MSSM sectors and the case in which there isn't such mixing.\\ In the no-mixing case we are able to keep 
fixed the mass gap between the LSP and the NLSP. We have found that if the NLSP is mostly a bino we can satisfy the
WMAP up to $10\%$ mass gap; if the NLSP is mostly a wino  we can satisfy the
WMAP up to $20\%$ mass gap.  \\In the mixing case we have we have obtained a constraint on $Q_{H_u}$. We have also found 
sizable regions in which we can satisfy the WMAP data for mass gaps which go from $5 \%$ to beyond $20 \%$.
We have studied some specific sets of parameters for the $5\%$ and $10\%$ mass gap regions, 
showing that relatively small changes in the mass gap
can produce very important changes in the area of the regions which satisfy the experimental constraints.\\
We have also showed, as an example that we still have the MSSM physics, the presence of a funnel
region, analogous to that in the MSSM, in some region of the parameter space.\\
So we can say that a model with an anomalous LSP can satisfy all the current experimental constraints,
can show a phenomenology similar to that expected from a MSSM LSP
and can be viable to explain the DM abundance without any arbitrary constraints on its parameters.


\begin{thebibliography}{999}
\bibitem{Marchesano:2007de}
  F.~Marchesano,
  Fortsch.\ Phys.\  {\bf 55} (2007) 491
  [arXiv:hep-th/0702094].

\bibitem{Blumenhagen:2005mu}
  R.~Blumenhagen, M.~Cvetic, P.~Langacker and G.~Shiu,
  Ann.\ Rev.\ Nucl.\ Part.\ Sci.\  {\bf 55} (2005) 71
  [arXiv:hep-th/0502005].

\bibitem{Lust:2004ks}
  D.~Lust,
  Class.\ Quant.\ Grav.\  {\bf 21} (2004) S1399
  [arXiv:hep-th/0401156].

\bibitem{Kiritsis:2003mc}
  E.~Kiritsis,
  Fortsch.\ Phys.\  {\bf 52} (2004) 200
  [Phys.\ Rept.\  {\bf 421} (2005\ ERRAT,429,121-122.2006) 105]
  [arXiv:hep-th/0310001].

\bibitem{Langacker:2008yv}
  P.~Langacker,
  Rev.\ Mod.\ Phys.\  {\bf 81} (2009) 1199
  [arXiv:0801.1345 [hep-ph]].


\bibitem{Anastasopoulos:2008jt}
  P.~Anastasopoulos, F.~Fucito, A.~Lionetto, G.~Pradisi, A.~Racioppi and Y.~S.~Stanev,
  Phys.\ Rev.\  D {\bf 78} (2008) 085014
  [arXiv:0804.1156 [hep-th]].

\bibitem{Fucito:2010dj}
  F.~Fucito, A.~Lionetto, A.~Racioppi and D.~R.~Pacifici,
  Phys.\ Rev.\  D {\bf 82} (2010) 115004
  [arXiv:1007.5443 [hep-ph]].

\bibitem{FLMR}
  F.~Fucito, A.~Lionetto, A.~Mammarella, A.~Racioppi
  The European Physichal Journal C, 2010 Vol. 69, pag. 455-465
  [arXiv:0811.1953v3 [hep-ph]].

\bibitem{FLM}
  F.~Fucito, A.~Lionetto, A.~Mammarella, A.~Racioppi
 to be published on Phys. Rev. D.
  [arXiv:arXiv:1105.4753 [hep-ph]].


\bibitem{Coriano:2010ww}
  C.~Coriano, M.~Guzzi and A.~Mariano,
  arXiv:1012.2420 [hep-ph].

\bibitem{Coriano:2010ws}
  C.~Coriano, M.~Guzzi and A.~Mariano,
  arXiv:1010.2010 [hep-ph].

\bibitem{Kors:2005uz}
  B.~Kors and P.~Nath,
  JHEP {\bf 0507} (2005) 069
  [arXiv:hep-ph/0503208].

\bibitem{Feldman:2006wd}
  D.~Feldman, B.~Kors and P.~Nath,
  Phys.\ Rev.\  D {\bf 75} (2007) 023503
  [arXiv:hep-ph/0610133].

\bibitem{Klein:1999im}
  M.~Klein,
  Nucl.\ Phys.\  B {\bf 569} (2000) 362
  [arXiv:hep-th/9910143].
\bibitem{reliccoann}
  K.~Griest and D.~Seckel,
  Phys.\ Rev.\  D {\bf 43} (1991) 3191.

\bibitem{Edsjo:2003us}
  J.~Edsjo, M.~Schelke, P.~Ullio and P.~Gondolo,
  JCAP {\bf 0304} (2003) 001
  [arXiv:hep-ph/0301106].

\bibitem{Gondolo}
  J.~Edsjo and P.~Gondolo,
  Phys.\ Rev.\  D {\bf 56} (1997) 1879
  [arXiv:hep-ph/9704361].

\bibitem{WMAP}
  E.~Komatsu et al,
  The Astrophysical Journal Supplement Series 192:18 (2011)

\bibitem{tesiandrea}
A.~Mammarella, ``Phenomenology of anomalous U(1):
Dark Matter and Asymmetries'' (2011), Ph.D. thesis
(in preparation)

\bibitem{DarkSUSY}
P. Gondolo, J. Edsj\"{o}, P. Ullio, L. Bergstr\"{o}m, M. Schelke and E.A. Baltz,
JCAP 07 (2004) 008 [arXiv:astro-ph/0406204]



\end{thebibliography}
\end{document}